\documentstyle[preprint,eqsecnum,aps]{revtex}

\begin{document}
\draft


\preprint{YITP-98-35, gr-qc/9806012}

\title{Black holes, brick walls and the Boulware state}
\author{Shinji Mukohyama and Werner Israel~\footnote{
Permanent address: 
{\it Canadian Institute for Advanced Research Cosmology Program,
Department of Physics and Astronomy, University of Victoria, Victoria
BC, Canada V8W 3P6}.
}}
\address{
Yukawa Institute for Theoretical Physics, 
Kyoto University \\
Kyoto 606-8502, Japan
}
\date{\today}

\maketitle


\begin{abstract} 

The brick-wall model seeks to explain the Bekenstein-Hawking entropy
as a wall-contribution to the thermal energy of ambient quantum fields 
raised to the Hawking temperature. Reservations have been expressed
concerning the self-consistency of this model. For example, it
predicts large thermal energy densities near the wall, producing a
substantial mass-correction and, presumably, a large gravitational
back-reaction. We re-examine this model and conclude that these
reservations are unfounded once the ground state---the Boulware
state---is correctly identified. We argue that the brick-wall model
and the Gibbons-Hawking instanton (which ascribes a topological origin 
to the Bekenstein-Hawking entropy) are mutually exclusive, alternative 
descriptions (complementary in the sense of Bohr) of the same
physics. 

\end{abstract}


\newpage


\section{Introduction}

The prescription 
%
\begin{equation}
 S_{BH} = \frac{1}{4}A/l_{pl}^2 \label{eqn:S_BH}
\end{equation}
for assigning a ``Bekenstein-Hawking entropy'' $S_{BH}$ to a black 
hole of surface area $A$ was first inferred in the mid-1970s from the 
formal similarities between black hole dynamics and 
thermodynamics~\cite{Bekenstein}, combined with Hawking's 
discovery~\cite{Hawking} that black hole radiate thermally with a 
characteristic (Hawking) temperature
%
\begin{equation}
 T_{H} = \hbar\kappa_0 /2\pi,\label{eqn:T_H}
\end{equation}
where $\kappa_0$ is the surface gravity. That $S_{BH}$ is a genuine 
thermodynamical entropy is given credence by the ``generalized second 
law''~\cite{Zureck&Thorne,GSL}, which states that the sum of $S_{BH}$ 
and the entropy of surrounding matter is non-decreasing in any 
(quasi-static) interaction.

Much work and discussion have been devoted to the problem of deriving 
and understanding the enigmatic formula (\ref{eqn:S_BH}) in a more 
direct and fundamental way.

A statistical derivation of (\ref{eqn:S_BH}) for stationary black 
holes, using analytic continuation to the Euclidean sector and 
imposing a Matsubara period $T_{H}^{-1}$ on Euclidean time, was 
developed by Gibbons and Hawking~\cite{Gibbons&Hawking} in 1977. 
According to their derivation, $S_{BH}$ emerges already at zero-loop 
order, as a contribution to the partition function from the 
extrinsic-curvature boundary terms which accompany the 
Einstein-Hilbert gravitational action. (Only the outer boundary at 
infinity actually contributes. There is no inner boundary, because 
the horizon is represented by a regular point in the Euclidean sector 
of a non-extremal black hole with the above periodic identification, 
corresponding to the Hartle-Hawking state for ambient quantum 
fields.)

Unfortunately, the Euclidean approach provides no insight into the 
dynamical origin of $S_{BH}$. On the contrary, it appears to suggest 
that one has to think of $S_{BH}$ as having a topological origin in 
some sense.

Interesting and suggestive, but still short of ideal from a physical 
point of view, are the interpretations which require an appeal to the 
past or future history of the black hole, or to ensembles of such 
histories; for instance, ``$\exp (S_{BH})$ represents the number of 
quantum-mechanically distinct ways in which the black hole could have 
been made''~\cite{Zureck&Thorne}, or those which relate $S_{BH}$ to 
the entropy of the evaporation products~\cite{EvaporationProd,York}.

In thermodynamics, entropy is a state function with a definite value 
at each moment of time (at least for quasi-stationary, 
near-equilibrium processes). Correspondingly, one would ideally like 
to have a direct physical understanding of $S_{BH}$ for a specific 
black hole at a given moment in terms of the dynamical degrees of 
freedom existing at that moment, together with an explanation of how 
it comes to have the simple universal form (\ref{eqn:S_BH}), 
independent of the nature and number of the fundamental fields and 
all details of the microphysics~\cite{Frolov&Novikov}.

An interpretation which holds promise of meeting these requirements is 
that $S_{BH}$ is entanglement entropy, associated with modes and 
correlations hidden from external observers by the presence of the 
horizon. If the black hole originates from a pure state, there is 
perfect correlation between the internal and external modes, and the 
entanglement entropy can therefore equally well be found by counting 
external modes. A program of this type was first clearly formulated 
by Bombelli, Koul, Lee and Sorkin~\cite{BKLS} in a 1986 paper which 
attracted little notice at the time. It was independently 
re-initiated by Srednicki~\cite{Srednicki} and by Frolov and 
Novikov~\cite{Frolov&Novikov} in 1993.

It turns out, remarkably, that the entanglement entropy is 
proportional to the area of the dividing wall. This is a robust 
result which holds not only for black hole 
horizons~\cite{Frolov&Novikov,MSK1998}, but also for cavities 
artificially cut out of Minkowski space~\cite{BKLS,Srednicki,MSK1997}. 
However, the coefficient of proportionality is formally infinite, 
corresponding to the fact that modes of arbitrarily small wavelength 
can exist close to the horizon or partition. (This, incidentally, 
yields the correct answer---infinity---for the entropy of a classical 
black hole.) But an infinitely sharp boundary violates quantum 
mechanics. Quantum fluctuations will prevent events closer to the 
horizon than about a Planck length $l_{pl}$ from being seen on the 
outside. Introducing an effective cutoff of this order reproduces the 
Bekenstein-Hawking formula (\ref{eqn:S_BH}) with a coefficient of the 
right order of magnitude.

The entanglement interpretation seems to be implicit in, and is 
certainly closely related to a pioneering calculation done by Gerard 
`tHooft~\cite{tHooft} in 1985. He considered the statistical 
thermodynamics of quantum fields in the Hartle-Hawking state (i.e. 
having the Hawking temperature $T_{H}$ at large radii) propagating on 
a fixed Schwarzschild background of mass $M$. To control divergences, 
`tHooft introduced a ``brick wall''---actually a static spherical 
mirror at which the fields are required to satisfy Dirichlet or 
Neumann boundary conditions---with radius a little larger than the 
gravitational radius $2M$. He found, in addition to the expected 
volume-dependent thermodynamical quantities describing hot fields in a 
nearly flat space, additional wall contributions proportional to the area. 
These contributions are, however, also proportional to $\alpha^{-2}$, 
where $\alpha$ is the proper altitude of the wall above the 
gravitational radius, and thus diverge in the limit $\alpha\to 0$. For 
a specific choice of $\alpha$ (which depends on the number of fields, 
etc., but is generally of order $l_{pl}$), `tHooft was able to 
recover the Bekenstein-Hawking formula (\ref{eqn:S_BH}) with the 
correct coefficient.

However, this calculation raises a number of questions which have 
caused many, including `tHooft himself, to have reservations about 
its validity and consistency. 

\begin{enumerate}
 \renewcommand{\labelenumi}{(\alph{enumi})}
 \item $S_{BH}$ is here obtained as a one-loop effect, originating 
 from thermal excitations of the quantum fields. Does this material 
 contribution to $S_{BH}$ have to be {\it added} to the zero-loop 
 Gibbons-Hawking contribution which arises from the gravitational 
 part of the action and already by itself accounts for the full value 
 of $S_{BH}$?~\cite{Liberati}
 \item The ambient quantum fields were assumed to be in the 
 Hartle-Hawking state. Their stress-energy should therefore be 
 bounded (of order $M^{-4}$ in Planck units) near the gravitational 
 radius, and negligibly small for large masses. However, `tHooft's 
 calculation assigns to them enormous (Planck-level) energy densities 
 near the wall. 
 \item The integrated field energy gives a wall contribution to the
mass 
 %
 \begin{equation}
  \Delta M = \frac{3}{8}M \label{eqn:DeltaM=3/8M}
 \end{equation}
 when $\alpha$ is adjusted to give the correct value of $S_{BH}$. This 
 suggests a substantial gravitational back-reaction~\cite{tHooft} and 
 that the assumption of a fixed geometrical background may be 
 inconsistent~\cite{Liberati,Belgiorno&Martellini,Susskind&Uglum}. 
\end{enumerate}

Our main purpose in this paper is to point out that these 
difficulties are only apparent and easy to resolve. The basic remark 
is that the {\it brick-wall model strictly interpreted does not 
represent a black hole}. It represents the exterior of a starlike 
object with a reflecting surface, compressed to nearly (but not 
quite) its gravitational radius. The ground state for quantum fields 
propagating around this star is not the Hartle-Hawking state but the 
Boulware state~\cite{Boulware}, corresponding to zero
temperature, which has a quite different behavior near the
gravitational radius.

The Boulware state in a static spherical geometry is defined as being
free of modes having positive frequency with respect to the
conventional Schwarzschild time $t$. At infinity it approaches the
Minkowski vacuum with zero stress-energy. At smaller radii, vacuum
polarization induces a non-vanishing stress-energy, which diverges
near a brick wall skirting the gravitational radius. The asymptotic
behavior is that of a thermal stress-energy---the characteristic
temperature is the local acceleration temperature for static observers, 
diverging near the gravitational radius---{\it but with the opposite
sign}. The Boulware state is energetically depressed below the vacuum.
(We shall reserve the word ``vacuum'' for a condition of zero
stress-energy; in general, it is not a quantum state.)

The quantum fields with temperature $T_H$ at infinity which 'tHooft
introduced into his brick wall model have a local temperature and a
thermal stress-energy which also diverge near the gravitational
radius. To obtain the total (gravitating) stress-energy, one must add
the contributions of the (Boulware) ground state and the thermal
excitations. The diverging parts of these contributions cancel. The
sum is bounded, of order $M^{-4}$ near the gravitational radius (hence 
small for large masses), and in the limit $\alpha\to 0$
indistinguishable from the Hartle-Hawking stress-energy.

Thus, it is perfectly legitimate to neglect back-reaction in the brick 
wall model. With the proper identification of the ground state,
problems (b) and (c) above resolve themselves.

The same basic observation resolves problem (a). In the Euclidean
sector of the brick-wall spacetime (with Matsubara period $T_H^{-1}$), 
it is not true that there is no inner boundary. The inner boundary is
the brick wall itself, and its boundary contribution to the Euclidean
action cancels that of the outer boundary at infinity. So the
Gibbons-Hawking zero-loop contribution is now zero. The ``geometrical
entropy'' of the Gibbons-Hawking ``instanton'' thus provides an
alternative, complementary description of $S_{BH}$, not a
supplementary contribution to the entropy of thermal excitations as
calculated from the brick wall model.

In Sec.~\ref{sec:B&HH-states} we summarize essential properties of the 
Boulware and Hartle-Hawking states that play a role in this paper. 
In Sec.~\ref{sec:BW-model} we sketch the physical essence of the
brick-wall model by using a particle description of quantum fields. 
A systematic treatment of the model is deferred to
Sec.~\ref{sec:BW-model2}, in which the results in
Sec.~\ref{sec:BW-model} are rigorously derived from the quantum field
theory in curved spacetimes. 
Sec.~\ref{sec:discussion} is devoted to summarize this paper. 
In Appendix~\ref{app:on-shell}, for completeness, we apply the
so-called on-shell method to the brick wall model and show that in the 
on-shell method we might miss some physical degrees of freedom. Hence, 
we do not adopt the on-shell method in the main body of this paper.


\section{The Boulware and Hartle-Hawking states}
	\label{sec:B&HH-states}

It is useful to begin by summarizing briefly the essential properties
of the quantum states that will play a role in our discussion.

In a curved spacetime there is no unique choice of time
coordinate. Different choices lead to different definitions of
positive-frequency modes and different ground states.

In any static spacetime with static (Killing) time parameter $t$, the
Boulware state $|B\rangle$ is the one annuled by the annihilation
operators $a_{Kill}$ associated with ``Killing modes''
(positive-frequency in $t$). In an asymptotically flat space,
$|B\rangle$ approaches the Minkowski vacuum at infinity.

In the spacetime of a stationary eternal black hole, the
Hartle-Hawking state $|HH\rangle$ is the one annuled by $a_{Krus}$,
the annihilation operators associated with ``Kruskal modes''
(positive-frequency in the Kruskal lightlike coordinates $U$,
$V$). This state appears empty of ``particles'' to free falling
observers at the horizon, and its stress-energy is bounded there (not
quite zero, because of irremovable vacuum polarization effects).

If, just for illustrative purpose, we consider a $(1+1)$-dimensional
spacetime, it is easy to give concrete form to these remarks. We
consider a spacetime with metric 
%
\begin{equation}
 ds^2 = -f(r)dt^2+\frac{dr^2}{f(r)},
	\label{eqn:2d-metric}
\end{equation}
and denote by $\kappa(r)$ the redshifted gravitational force, i.e.,
the upward acceleration $a(r)$ of a stationary test-particle reduced
by the redshift factor $f^{1/2}(r)$, so that
$\kappa(r)=\frac{1}{2}f'(r)$. A horizon is characterized by $r=r_0$,
$f(r_0)=0$, and its surface gravity defined by
$\kappa_0=\frac{1}{2}f'(r_0)$.

Quantum effects induce an effective quantum stress-energy $T_{ab}$
($a,b,\cdots =r,t$) in the background geometry
(\ref{eqn:2d-metric}). If we assume no net energy flux 
($T^r_t=0$)---thus excluding the Unruh state---$T_{ab}$ is completely
specified by a quantum energy density $\rho=-T^t_t$ and pressure
$P=T^r_r$. These are completely determined (up to a boundary
condition) by the conservation law $T^b_{a;b}=0$ and the trace
anomaly, which is 
%
\begin{equation}
 T^a_a = \frac{\hbar}{24\pi}R
\end{equation}
for a massless scalar field, with $R=-f''(r)$ for the metric
(\ref{eqn:2d-metric}). Integration gives 
%
\begin{equation}
 f(r)P(r) = -\frac{\hbar}{24\pi}(\kappa^2(r) + const.).
	\label{eqn:2d-fP}
\end{equation}
Different choices of the constant of integration correspond to
different boundary conditions, i.e., to different quantum states.

For the Hartle-Hawking state, we require $P$ and $\rho$ to be bounded
at the horizon $r=r_0$, giving 
%
\begin{eqnarray}
 P_{HH} & = & \frac{\hbar}{24\pi}\frac{\kappa_0^2-\kappa^2(r)}{f(r)},
	\nonumber\\
 \rho_{HH} & = & P_{HH} + \frac{\hbar}{24\pi}f''(r).
	\label{eqn:HH-P-rho}
\end{eqnarray}
When $r\to\infty$ this reduces to (setting $f(r)\to 1$)
%
\begin{eqnarray}
 \rho_{HH} & \simeq & P_{HH} = \frac{\pi}{6\hbar}T_H^2,
	\nonumber\\
 T_H & = & \hbar\kappa_0/2\pi,
\end{eqnarray}
which is appropriate for one-dimensional scalar radiation at the
Hawking temperature $T_H$.

For the Boulware state, the boundary condition is $P=\rho=0$ when
$r=\infty$. The integration constant in (\ref{eqn:2d-fP}) must vanish, 
and we find 
%
\begin{eqnarray}
 P_{B} & = & -\frac{\hbar}{24\pi}\frac{\kappa^2(r)}{f(r)},
	\nonumber\\
 \rho_{B} & = & P_{B} + \frac{\hbar}{24\pi}f''(r).
	\label{eqn:B-P-rho}
\end{eqnarray}
If a horizon were present, $\rho_B$ and $P_B$ would diverge there to
$-\infty$.

For the difference of these two stress tensors,
%
\begin{eqnarray}
 \Delta T_a^b = (T_a^b)_{HH} - (T_a^b)_{B},\label{eqn:DeltaTab}
\end{eqnarray}
(\ref{eqn:HH-P-rho}) and (\ref{eqn:B-P-rho}) give the exactly thermal
form 
%
\begin{eqnarray}
 \Delta P = \Delta\rho = \frac{\pi}{6\hbar}T^2(r),
	\label{eqn:2d-DP-Drho}
\end{eqnarray}
where $T(r)=T_H/\sqrt{f(r)}$ is the local temperature in the
Hartle-Hawking state. We recall that thermal equilibrium in any static 
gravitational field requires the local temperature $T$ to rise with
depth in accordance with Tolman's law~\cite{Tolman}
%
\begin{eqnarray}
 T\sqrt{-g_{00}} = const. \label{eqn:Tolman}
\end{eqnarray}

We have found, for this $(1+1)$-dimensional example, that the
Hartle-Hawking state is thermally excited above the zero-temperature
(Boulware) ground state to a local temperature $T(r)$ which grows
without bound near the horizon. Nevertheless, it is the Hartle-Hawking 
state which best approximates what a gravitational theorist would call 
a ``vacuum'' at the horizon.

These remarks remain at least qualitatively valid in
$(3+1)$-dimensions, with obvious changes arising from the 
dimensionality. In particular, the $(3+1)$-dimensional analogue of
(\ref{eqn:2d-DP-Drho}) for a massless scalar field, 
%
\begin{eqnarray}
 3\Delta P \simeq \Delta\rho 
	\simeq \frac{\pi^2}{30\hbar^3}T^4(r),\label{eqn:DeltaP-Deltarho}
\end{eqnarray}
holds to a very good approximation, both far from the black hole and
near the horizon. In the intermediate zone there are deviations, 
but they always remain bounded~\cite{DP-Drho}, and will not affect our 
considerations.


\section{Brick-wall model}
	\label{sec:BW-model}
	
We shall briefly sketch the physical essence of the brick-wall 
model. (A systematic treatment is deferred to 
Sec.~\ref{sec:BW-model2})

We wish to study the thermodynamics of hot quantum fields confined to 
the outside of a spherical star with a perfectly reflecting surface 
whose radius $r_{1}$ is a little larger than its gravitational radius 
$r_{0}$. To keep the total field energy bounded, we suppose the 
system enclosed in a spherical container of radius $L\gg r_{1}$.

It will be sufficiently general to assume for the geometry outside 
the star a spherical background metric of the form 
%
\begin{equation}
 ds^2 = -f(r)dt^2+\frac{dr^2}{f(r)}+r^2d\Omega^2.
 	\label{eqn:4d-metric}
\end{equation}
This includes as special cases the Schwarzschild, 
Reissner-Nordstr{\"o}m and de Sitter geometries, or any combination of 
these.

Into this space we introduce a collection of quantum fields, raised to 
some temperature $T_{\infty}$ at large distances, and in thermal 
equilibrium. The local temperature $T(r)$ is then given by Tolman's 
law (\ref{eqn:Tolman}), 
%
\begin{equation}
 T(r) = T_{\infty}f^{-1/2}\label{eqn:T(r)-Tinfty}
\end{equation}
and becomes very large when $r\to r_{1}=r_{0}+\Delta r$. We shall 
presently identify $T_{\infty}$ with the Hawking temperature $T_{H}$ 
of the horizon $r=r_{0}$ of the exterior metric 
(\ref{eqn:4d-metric}), continued (illegitimately) into the internal
domain $r<r_{1}$.

Characteristic wavelengths $\lambda$ of this radiation are small 
compared to other relevant length-scales (curvature, size of
container) in the regions of interest to us. Near the star's surface, 
%
\begin{equation}
 \lambda \sim \hbar/T = f^{1/2}\hbar/T_{\infty}\ll r_{0}.\nonumber
\end{equation}
Elsewhere in the large container, at large distances from the star, 
%
\begin{equation}
 f\simeq 1, \quad
 \lambda\simeq\hbar/T_{\infty}\sim r_{0}\ll L.\nonumber
\end{equation}
Therefore, a particle description should be a good approximation to 
the statistical thermodynamics of the fields (Equivalently, one can 
arrive at this conclusion by considering the WKB solution to the wave 
equation, cf. `tHooft~\cite{tHooft} and 
Sec.~\ref{sec:BW-model2})

For particles of rest-mass $m$, energy $E$, $3$-momentum $p$ and 
$3$-velocity $v$ as viewed by a local stationary observer, the energy 
density $\rho$, pressure $P$ and entropy density $s$ are given by the 
standard expressions 
%
\begin{eqnarray}
 \rho & = & {\cal N}\int_{0}^{\infty}
 	\frac{E}{e^{\beta E}-\epsilon}\frac{4\pi p^2dp}{h^3},
 	\nonumber\\
 P & = & \frac{{\cal N}}{3}\int_{0}^{\infty}
 	\frac{vp}{e^{\beta E}-\epsilon}\frac{4\pi p^2dp}{h^3},
 	\nonumber\\
 s & = & \beta (\rho +P).\label{eqn:S-P-rho}
\end{eqnarray}
Here, as usual, 
%
\begin{equation}
 E^2-p^2 =m^2, \quad v=p/E, \quad \beta =T^{-1};\nonumber
\end{equation}
$\epsilon$ is $+1$ for bosons and $-1$ for fermions and the factor 
${\cal N}$ takes care of helicities and the number of particle 
species. The total entropy is given by the integral 
%
\begin{equation}
 S = \int_{r_{1}}^{L}s(r)4\pi r^2dr/\sqrt{f},
 	\label{eqn:Stot}
\end{equation}
where we have taken account of the proper volume element as given by 
the metric (\ref{eqn:4d-metric}). The factor $f^{-1/2}$ does not,
however, appear in the integral for the gravitational mass of the
thermal excitations~\cite{MTW} (It is canceled, roughly speaking, by 
negative gravitational potential energy):
%
\begin{equation}
 \Delta M_{therm} = \int_{r_{1}}^{L}\rho(r)4\pi r^2dr.
	\label{eqn:Mtherm}
\end{equation}

The integrals (\ref{eqn:Stot}) and (\ref{eqn:Mtherm}) are dominated by 
two contributions for large container radius $L$ and for small 
$\Delta r=r_{1}-r_{0}$:
\begin{enumerate}
 \renewcommand{\labelenumi}{(\alph{enumi})}
 \item A volume term, proportional to $\frac{4}{3}\pi L^3$, 
 representing the entropy and mass-energy of a homogeneous quantum 
 gas in a flat space (since $f\simeq 1$ almost everywhere in the 
 container if $L/r_{0}\to\infty$) at a uniform temperature 
 $T_{\infty}$. This is the result that would have been expected, and 
 we do not need to consider it in detail. 
 \item Of more interest is the contribution of gas near the inner 
 wall $r=r_{1}$, which we now proceed to study further. We shall find 
 that it is proportional to the wall area, and diverging like 
 $(\Delta r)^{-1}$ when $\Delta r\to 0$.
\end{enumerate}

Because of the high local temperatures $T$ near the wall for small 
$\Delta r$, we may insert the ultrarelativistic approximations 
%
\[
 E\gg m,\quad p\simeq E,\quad v\simeq 1
\]
into the integrals (\ref{eqn:S-P-rho}). This gives
%
\begin{equation}
 P\simeq\frac{1}{3}\rho
 	\simeq\frac{{\cal N}}{6\pi^2}T^4\int_{0}^{\infty}
 	\frac{x^3dx}{e^x-\epsilon}
\end{equation}
in Planck units ($h=2\pi\hbar =2\pi$). The purely numerical integral 
has the value $3!$ multiplied by $1$, $\pi^4/90$ and 
$\frac{7}{8}\pi^4/90$ for $\epsilon=0,1$ and $1$ respectively, and we 
shall adopt $3!$, absorbing any small discrepancy into ${\cal N}$. 
Then, from (\ref{eqn:S-P-rho}), 
%
\begin{equation}
 \rho=\frac{3{\cal N}}{\pi^2}T^4,\quad s=\frac{4{\cal N}}{\pi^2}T^3
 	\label{eqn:rho-T4}
\end{equation}
in terms of the local temperature $T$ given by (\ref{eqn:T(r)-Tinfty}).

Substituting (\ref{eqn:rho-T4}) into (\ref{eqn:Stot}) gives for the 
wall contribution to the total entropy
%
\begin{equation}
 S_{wall} = \frac{4{\cal N}}{\pi^2}4\pi r_{1}^2 T_{\infty}^3
 	\int_{r_{1}}^{r_{1}+\delta}\frac{dr}{f^2(r)},
 	\label{eqn:Swall}
\end{equation}
where $\delta$ is an arbitrary small length subject to 
$\Delta r\ll\delta\ll r_{1}$. It is useful to express this result in 
terms of the proper altitude $\alpha$ of the inner wall above the 
horizon $r=r_{0}$ of the exterior geometry (\ref{eqn:4d-metric}). 
(Since (\ref{eqn:4d-metric}) only applies for $r>r_{1}$, the physical 
space does not, of course, contain any horizon.) We assume that $f(r)$ 
has a (simple) zero for $r=r_{0}$, so we can write
%
\begin{equation}
 f(r) \simeq 2\kappa_{0} (r-r_{0}),\quad
 \kappa_{0}=\frac{1}{2}f'(r_{0})\ne 0 \quad
 (r\to r_{0}),
\end{equation}
where $\kappa_{0}$ is the surface gravity. Then 
%
\begin{equation}
 \alpha = \int_{r_{0}}^{r_{1}}f^{-1/2}dr\quad\Rightarrow\quad
 	\Delta r = \frac{1}{2}\kappa_0\alpha^2,
\end{equation}
and (\ref{eqn:Swall}) can be written 
%
\begin{equation}
 S_{wall} = \frac{{\cal N}}{90\pi\alpha^2}
 	\left(\frac{T_{\infty}}{\kappa_0/2\pi}\right)^3
	\frac{1}{4}A \label{eqn:Swall2}
\end{equation}
in Planck units, where $A=4\pi r_{1}^2$ is the wall area.

Similarly, we find from (\ref{eqn:Mtherm}) and (\ref{eqn:rho-T4}) that 
thermal excitations near the wall contribute 
%
\begin{equation}
 \Delta M_{them,wall} = \frac{{\cal N}}{480\pi\alpha^2}
 	\left(\frac{T_{\infty}}{\kappa_0/2\pi}\right)^3AT_{\infty}
 	\label{eqn:M-thermw-wall}
\end{equation}
to the gravitational mass of the system.

The wall contribution to the free energy 
%
\begin{equation}
 F=\Delta M-T_{\infty}S \label{eqn:Gibbs-Duhem}
\end{equation}
is 
%
\begin{equation}
 F_{wall}= -\frac{{\cal N}}{1440\pi\alpha^2}
 	\left(\frac{T_{\infty}}{\kappa_0/2\pi}\right)^3AT_{\infty}.
 	\label{eqn:Fwall}
\end{equation}
The entropy is recoverable from the free energy by the standard 
prescription
%
\begin{equation}
 S_{wall} = -\partial F_{wall}/\partial T_{\infty}.\label{eqn:Gibbs}
\end{equation}
(Observe that this is an ``off-shell'' prescription~\cite{on-shell}: 
the geometrical quantities $A$, $\alpha$ and, in particular, the 
surface gravity $\kappa_{0}$ are kept fixed when the temperature is 
varied in (\ref{eqn:Fwall}).)

Following `tHooft~\cite{tHooft}, we now introduce a crude cutoff to 
allow for quantum-gravity fluctuations by fixing the wall altitude 
$\alpha$ so that 
%
\begin{equation}
 S_{wall} = S_{BH},\quad \mbox{when}\quad T_{\infty} = T_{H},
 	\label{eqn:Swall=SBH}
\end{equation}
where the Bekenstein-Hawking entropy $S_{BH}$ and Hawking temperature 
$T_{H}$ are defined to be the {\it purely geometrical} quantities 
defined by (\ref{eqn:S_BH}) and (\ref{eqn:T_H}) in terms of the wall's 
area $A$ and redshifted acceleration ($=$ surface gravity) $\kappa_0$. 
From (\ref{eqn:Swall=SBH}) and (\ref{eqn:Swall2}), restoring 
conventional units for a moment, we find 
%
\begin{equation}
 \alpha = l_{pl}\sqrt{{\cal N}/90\pi},\label{eqn:normalization}
\end{equation}
so that $\alpha$ is very near the Planck length if the effective 
number ${\cal N}$ of basic quantum fields in nature is on the order 
of $300$.

It is significant and crucial that the normalization 
(\ref{eqn:normalization}) is {\it universal}, depending only on 
fundamental physics, and independent of the mechanical and 
geometrical characteristics of the system.

With $\alpha$ fixed by (\ref{eqn:normalization}), the wall's free 
energy (\ref{eqn:Gibbs-Duhem}) becomes 
%
\begin{equation}
 F_{wall} = -\frac{1}{16}\left(\frac{T_{\infty}}{T_{H}}\right)^3
 	AT_{\infty}.\label{eqn:Fwall2}
\end{equation}
This ``off-shell'' formula expresses $F_{wall}$ in terms of three 
independent variables: the temperature $T_{\infty}$ and the 
geometrical characteristics $A$ and $T_{H}$. From (\ref{eqn:Fwall2}) 
we can obtain the wall entropy either from the thermodynamical Gibbs 
relation (\ref{eqn:Gibbs}) (with $T_{\infty}$ set equal to $T_{H}$ 
{\it after} differentiation), or from the Gibbs-Duhem formula 
(\ref{eqn:Gibbs-Duhem}) which is equivalent to the 
statistical-mechanical definition $S=-{\bf Tr}(\rho\ln\rho)$. Thus the 
distinction~\cite{on-shell} between ``thermodynamical'' and 
``statistical'' entropies disappears in this formulation, because the 
geometrical and thermal variables are kept independent.

The wall's thermal mass-energy is given ``on-shell'' 
($T_{\infty}=T_{H}$) by 
%
\begin{equation}
 \Delta M_{therm,wall} = \frac{3}{16}AT_{H}
\end{equation}
according to (\ref{eqn:M-thermw-wall}) and (\ref{eqn:normalization}). 
For a wall skirting a Schwarzschild horizon, so that 
$T_{H}=(8\pi M)^{-1}$, this reduces to `tHooft's result 
(\ref{eqn:DeltaM=3/8M}).

As already noted, thermal energy is not the only source of the wall's 
mass. Quantum fields outside the wall have as their ground state the 
Boulware state, which has a negative energy density growing to Planck 
levels near the wall. On shell, this very nearly cancels the thermal 
energy density (\ref{eqn:rho-T4}); their sum is, in fact, the 
Hartle-Hawking value (cf. (\ref{eqn:DeltaTab}) and 
(\ref{eqn:DeltaP-Deltarho})):  
%
\begin{equation}
 (T_{\mu}^{\nu})_{therm,T_{\infty}=T_{H}} + (T_{\mu}^{\nu})_{B}
 = (T_{\mu}^{\nu})_{HH},
\end{equation}
which remains bounded near horizons, and integrates virtually to zero 
for a very thin layer near the wall. The total gravitational mass of 
the wall is thus, from (\ref{eqn:M-thermw-wall}) and 
(\ref{eqn:normalization}),
%
\begin{eqnarray}
 (\Delta M)_{wall} & = & (\Delta M)_{therm,wall} + (\Delta M)_{B,wall}
 	\nonumber\\
 	& = & \frac{3}{16}AT_{H}\left( (T_{\infty}/T_{H})^4-1\right),
\end{eqnarray}
which vanishes on shell. For a central mass which is large in Planck 
units, there is no appreciable back-reaction of material near the wall 
on the background geometry (\ref{eqn:4d-metric}).

We may conclude that many earlier 
concerns~\cite{tHooft,Liberati,Belgiorno&Martellini} were 
unnecessary: `tHooft's brick wall model does provide a perfectly 
self-consistent description of a configuration which is 
indistinguishable from a black hole to outside observers, and which 
accounts for the Bekenstein-Hawking entropy purely as thermal entropy 
of quantum fields at the Hawking temperature (i.e. in the 
Hartle-Hawking state), providing one accepts the ad hoc but plausible 
ansatz (\ref{eqn:normalization}) for a Planck-length cutoff near the 
horizon.

The model does, however, present us with a feature which is 
theoretically possible but appears strange and counterintuitive from a 
gravitational theorist's point of view. Although the wall is 
insubstantial (just like a horizon)---i.e., space there is 
practically a vacuum and the local curvature low---it is 
nevertheless the repository of all of the Bekenstein-Hawking entropy 
in the model.

It has been argued~\cite{Frolov&Novikov} that this is just what might 
be expected of black hole entropy in the entanglement picture. 
Entanglement will arise from virtual pair-creation in which one 
partner is ``invisible'' and the other ``visible'' (although only 
temporarily---nearly all get reflected back off the potential barrier). 
Such virtual pairs are all created very near the horizon. Thus, on
this picture, the entanglement entropy (and its divergence) arises
almost entirely from the strong correlation between nearby field
variables on the two sides of the partition, an effect already present 
in flat space~\cite{Callan&Wilczek}.

An alternative (but not necessarily incompatible) possibility is that
the concentration of entropy at the wall is an artifact of the model
or of the choice of Fock representation (based on a static observer's
definition of positive frequency). The boundary condition of perfect
reflectivity at the wall has no black hole counterpart. Moreover, one
may well suspect that localization of entanglement entropy is not an
entirely well-defined concept~\cite{Mukohyama} or invariant under
changes of the Fock representation.


\section{The brick wall model revisited}
	\label{sec:BW-model2}

In the previous section, we have investigated the statistical 
mechanics of quantum fields in the region $r_{1}<r<L$ of the spherical 
background (\ref{eqn:4d-metric}) with the Dirichlet boundary condition 
at the boundaries. By using the particle description with the local 
temperature given by the Tolman's law, we have obtained the inner-wall 
contributions of the fields to entropy and thermal energy. When the 
former is set to be equal to the black hole entropy by 
fixing the cutoff $\alpha$ as (\ref{eqn:normalization}), the 
later becomes comparable with the mass of the background geometry. 
After that, it has been shown that at the Hawking temperature the wall 
contribution to the thermal energy is exactly canceled by the negative 
energy of the Boulware state, assuming implicitly that the ground state 
of the model is the Boulware state and that the  gravitational energy 
appearing in the Einstein equation is a sum of the renormalized energy 
of the Boulware state and the thermal energy of the fields.

In this section we shall show that these implicit assumptions do 
hold. In the following arguments it will also become clear how the 
local description used in the previous section is derived from the 
quantum field theory in curved spacetime, which is globally defined.

For simplicity, we consider a real scalar field described by the 
action 
%
\begin{equation}
 I = -\frac{1}{2}\int d^4x\sqrt{-g}\left[
 	g^{\mu\nu}\partial_{\mu}\phi\partial_{\nu}\phi
 	+ m_{\phi}^2\phi^2\right]. 
\end{equation}
On the background given by (\ref{eqn:4d-metric}), the action is reduced 
to 
%
\begin{equation}
 I =  \int dt L,
\end{equation}
with the Lagrangian $L$ given by
%
\begin{equation}
 L = - \frac{1}{2}\int d^3x r^2\sqrt{\Omega}\left[
 	- \frac{1}{f}(\partial_{t}\phi)^2 + f(\partial_{r}\phi)^2
 	+ \frac{1}{r^2}\Omega^{ij}\partial_{i}\phi\partial_{j}\phi
 	+ m_{\phi}\phi^2\right]. 
\end{equation}
Here $x^i$ ($i=1,2$) are coordinates on the $2$-sphere. 
In order to make our system finite let us suppose that two 
mirror-like boundaries are placed at $r=r_1$ and $r=L$ ($r_1<L$), 
respectively, and investigate the scalar field in the region between the 
two boundaries. In the following arguments we quantize the 
scalar field with respect to the Killing time $t$. Hence, the ground 
state obtained below is the Boulware state.
After the quantization, we investigate the statistical mechanics of the 
scalar field in the Boulware state. It will be shown that the resulting 
statistical mechanics is equivalent to the brick wall model.

Now let us proceed to the quantization procedure. First, the momentum 
conjugate to $\phi(r,x^i)$ is 
%
\begin{equation}
 \pi(r,x^i) = \frac{r^2\sqrt{\Omega}}{f}\partial_{t}\phi,
\end{equation}
and the Hamiltonian is given by
%
\begin{equation}
 H = \frac{1}{2}\int d^3x\left[\frac{f}{r^2\sqrt{\Omega}}\pi^2
 	+ r^2\sqrt{\Omega}f\left(\partial_{r}\phi\right)^2
 	+ \sqrt{\Omega}\Omega^{ij}\partial_{i}\phi\partial_{j}\phi
 	+ r^2\sqrt{\Omega}m_{\phi}^2\phi^2\right].
 	\label{eqn:Hamiltonian}
\end{equation}
Next, promote the field $\phi$ to an operator and expand it as
%
\begin{equation}
 \phi(r,x^i) = \sum_{nlm}\frac{1}{\sqrt{2\omega_{nl}}}\left[
 	a_{nlm}\varphi_{nl}(r)Y_{lm}(x^i)e^{-i\omega_{nl}t}
 	+ a^{\dagger}_{nlm}\varphi_{nl}(r)Y_{lm}(x^i)e^{i\omega_{nl}t}
 	\right],
\end{equation}
where $Y_{lm}(x^i)$ are real spherical harmonics defined by
%
\begin{eqnarray*}
 \frac{1}{\sqrt{\Omega}}\partial_{i}\left(\sqrt{\Omega}\Omega^{ij}
 	\partial_{j}Y_{lm}\right) + l(l+1)Y_{lm} & = & 0,\\
 \int Y_{lm}(x^i)Y_{l'm'}(x^i)\sqrt{\Omega(x^i)}d^2x
 	& = & \delta_{ll'}\delta_{mm'},
\end{eqnarray*}
and $\{\varphi_{nl}(r)\}$ ($n=1,2,\cdots$) is a set of real functions 
defined below, which is complete with respect to the space of 
$L_{2}$-functions on the interval $r_{1}\leq r\leq L$ for 
each $l$. The positive constant $\omega_{nl}$ is defined as the 
corresponding eigenvalue. 
%
\begin{eqnarray}
 \frac{1}{r^2}\partial_{r}\left( r^2f\partial_{r}\varphi_{nl}\right) 
 	- \left[ \frac{l(l+1)}{r^2}+m_{\phi}^2\right]\varphi_{nl}
 	+ \frac{\omega_{nl}^2}{f}\varphi_{nl} & = & 0,
 	\label{eqn:field-eq}\\
 \varphi_{nl}(r_{1})=\varphi_{nl}(L) & = & 0,\nonumber\\
 \int_{r_{1}}^{L}\varphi_{nl}(r)\varphi_{n'l}(r)\frac{r^2}{f(r)}dr
 	& = & \delta_{nn'}.\nonumber
\end{eqnarray}
The corresponding expansion of the operator $\pi(r,x^i)$ is then:
%
\begin{equation}
 \pi(r,x^i) = -i\frac{r^2\sqrt{\Omega(x^i)}}{f(r)}\sum_{nlm}
 	\sqrt{\frac{\omega_{nl}}{2}}\left[
 	a_{nlm}\varphi_{nl}(r)Y_{lm}(x^i)e^{-i\omega_{n}t}
 	- a^{\dagger}_{nlm}\varphi_{nl}(r)Y_{lm}(x^i)e^{i\omega_{nl}t}
 	\right]. 
\end{equation}
Hence, the usual equal-time commutation relation 
%
\begin{equation}
 \left[\phi(r,x^i),\pi(r',{x'}^i)\right] = 
 	i\delta(r-r')\delta^2(x^i-{x'}^i)
\end{equation}
becomes 
%
\begin{eqnarray}
 \left[ a_{nlm},a^{\dagger}_{n'l'm'}\right] & = & 
 	\delta_{nn'}\delta_{ll'}\delta_{mm'},\nonumber\\
 \left[ a_{nlm},a_{n'l'm'}\right] & = & 0,\nonumber\\
 \left[ a^{\dagger}_{nlm},a^{\dagger}_{n'l'm'}\right] & = & 0.
\end{eqnarray}
The normal-ordered Hamiltonian is given by
%
\begin{equation}
 :H: = \sum_{nlm}w_{nl}a^{\dagger}_{nlm}a_{nlm}.\label{eqn::H:}
\end{equation}
Thus, the Boulware state $|B\rangle$, which is defined by
%
\begin{equation}
 a_{nlm}|B\rangle = 0
\end{equation}
for ${}^{\forall}(n,l,m)$, is an eigenstate of the normal-ordered 
Hamiltonian with the eigenvalue zero. The Hilbert space of all quantum 
states of the scalar field is constructed as a symmetric Fock space 
on the Boulware state, and the complete basis 
$\{|\{N_{nlm}\}\rangle\}$ $(N_{nlm}=0,1,2,\cdots)$ is defined by
%
\begin{equation}
 |\{N_{nlm}\}\rangle = \prod_{nlm}\frac{1}{\sqrt{N_{nlm}!}}
 	\left( a^{\dagger}_{nlm}\right)^{N_{nlm}}|B\rangle,
\end{equation}
and each member of the basis is an eigenstate of the normal-ordered 
Hamiltonian:
%
\begin{equation}
 :H: |\{N_{nlm}\}\rangle = \left(\sum_{nlm}\omega_{nl}N_{nlm}\right) 
 	|\{N_{nlm}\}\rangle.
\end{equation}

Now we shall investigate the statistical mechanics of the quantized 
scalar field. The free energy $F$ is given by
%
\begin{equation}
 e^{-\beta_{\infty} F} \equiv
 	{\bf Tr}\left[ e^{-\beta_{\infty} :H:}\right]
	 = \prod_{nlm}\frac{1}{1-e^{-\beta_{\infty}\omega_{nl}}},
\end{equation}
where $\beta_{\infty}=T_{\infty}^{-1}$ is inverse temperature. 
For explicit calculation of the free energy we adopt the WKB 
approximation. First, we rewrite the mode function $\varphi_{nl}(r)$ 
as 
%
\begin{equation}
 \varphi_{nl}(r) = \psi_{nl}(r)e^{-ikr},
\end{equation}
and suppose that the prefactor $\psi_{nl}(r)$ varies very slowly: 
%
\begin{equation}
 \left|\frac{\partial_{r}\psi_{nl}}{\psi_{nl}}\right| \ll |k|,\qquad
 \left|\frac{\partial_{r}^2\psi_{nl}}{\psi_{nl}}\right| \ll |k|^2.
 	\label{eqn:WKB-cond1}
\end{equation}
Thence, assuming that
%
\begin{equation}
 \left|\frac{\partial_{r}(r^2f)}{r^2f}\right| \ll |k|,
 	\label{eqn:WKB-cond2}
\end{equation}
the field equation (\ref{eqn:field-eq}) of the mode function is 
reduced to 
%
\begin{equation}
 k^2 = k^2(l,\omega_{nl}) \equiv
 	\frac{1}{f}\left[\frac{\omega_{nl}^2}{f} 
 	- \frac{l(l+1)}{r^2} - m_{\phi}^2\right]. 
\end{equation}
Here we mention that the slowly varying condition 
(\ref{eqn:WKB-cond1}) can be derived from the condition 
(\ref{eqn:WKB-cond2}) and viceversa.
The number of modes with frequency less than $\omega$ is given
approximately by 
%
\begin{equation}
 \tilde{g}(\omega) = \int \nu({l,\omega}) (2l+1) dl,
\end{equation}
where $\nu(l,\omega)$ is the number of nodes in the mode with 
$(l,\omega)$:
%
\begin{equation}
 \nu(l,\omega) = \frac{1}{\pi}\int_{r_{1}}^{L} 
 	\sqrt{k^2(l,\omega)}dr.
\end{equation}
Here it is understood that the integration with respect to $r$ and 
$l$ is taken over those values which satisfy $r_{1}\le r\le L$ 
and $k^2(l,\omega)\geq 0$. Thus, when 
%
\[
 \left|\frac{\partial_{r}(r^2f)}{r^2f}\right| \ll 
 \frac{1}{f\beta_{\infty}}
\]
is satisfied, the free energy is given approximately by
%
\begin{equation}
 F \simeq \frac{1}{\beta_{\infty}}\int_{0}^{\infty} 
 	\ln\left(1-e^{-\beta_{\infty}\omega}\right)
 	\frac{d\tilde{g}(\omega)}{d\omega} d\omega 
 	= \int_{r_{1}}^{L}\tilde{F}(r) 4\pi r^2dr,
\end{equation}
where the `free energy density' $\tilde{F}(r)$ is defined by
%
\begin{equation}
 \tilde{F}(r) \equiv 
 	\frac{1}{\beta (r)}\int_{0}^{\infty}
 	\ln\left( 1-e^{-\beta (r) E}\right)
 	\frac{4\pi p^2dp}{(2\pi )^3}.
\end{equation}
Here the `local inverse temperature' $\beta (r)$ is defined by the 
Tolman's law
%
\begin{equation}
 \beta (r) = f^{1/2}(r)\beta_{\infty},
\end{equation}
and $E$ is defined by $E=\sqrt{p^2+m_{\phi}^2}$. Hence the total energy 
$U$ (equal to $\Delta M_{therm}$ given by (\ref{eqn:Mtherm})) 
and entropy $S$ are calculated as 
%
\begin{eqnarray}
 U & \equiv & {\bf Tr}\left[e^{\beta_{\infty}(F-:H:)}:H:\right]
 	= \frac{\partial}{\partial\beta_{\infty}}(\beta_{\infty} F) 
 	= \int_{r_{1}}^{L}\rho(r) 4\pi r^2dr,
 	\label{eqn:energy}\\
 S & \equiv & -{\bf Tr}\left[ e^{\beta_{\infty}(F-:H:)}
 	\ln e^{\beta_{\infty}(F-:H:)}\right]
 	= \beta_{\infty}^2\frac{\partial}{\partial\beta_{\infty}}F
 	= \int_{r_{1}}^{L}s(r) 4\pi r^2dr/\sqrt{f(r)},
 	\label{eqn:entropy}
\end{eqnarray}
where the `density' $\rho (r)$ and the `entropy density' $s(r)$ are 
defined by
%
\begin{eqnarray}
 \rho(r) & \equiv & \frac{\partial}{\partial\beta(r)}
 	(\beta(r) \tilde{F}(r)) 
 	= \int_{0}^{\infty}\frac{E}{e^{\beta(r)E}-1}
 	\frac{4\pi p^2dp}{(2\pi )^3}, \nonumber\\
 s(r) & \equiv & \beta^2(r)\frac{\partial}{\partial\beta(r)}
 	\tilde{F}(r)
  	= \beta(r)\left(\rho(r)+P(r)\right),\nonumber\\
\end{eqnarray}
where the `pressure' $P(r)$ is defined by~\footnote{
To obtain the last expression of $P(r)$ we performed an integration by 
parts. 
}
%
\begin{equation}
 P(r) \equiv -\tilde{F}(r)
 	= \frac{1}{3}\int_{0}^{\infty}\frac{p^2/E}{e^{\beta(r)E}-1}
 	\frac{4\pi p^2dp}{(2\pi )^3}.
\end{equation}
These expressions are exactly same as expressions (\ref{eqn:S-P-rho}) 
for the local quantities in the statistical mechanics of gas of 
particles.

Thus, we have shown that the local description of the statistical 
mechanics used in Sec.~\ref{sec:BW-model} is equivalent to that of the 
quantized field in the curved background, which is defined globally, 
and whose ground state is the Boulware state.

The stress energy tensor of the minimally coupled scalar field is given 
by 
%
\begin{equation}
 T_{\mu\nu} = 
 	-\frac{2}{\sqrt{-g}}\frac{\delta I}{\delta g^{\mu\nu}}
 = \partial_{\mu}\phi\partial_{\nu}\phi 
	-\frac{1}{2}g_{\mu\nu}\left(g^{\rho\sigma}
	\partial_{\rho}\phi\partial_{\sigma}\phi
	+m_{\phi}^2\phi^2\right).
\end{equation}
In particular, the $(tt)$-component is 
%
\begin{equation}
 T^t_t = -\frac{1}{2}\left[\frac{1}{f}(\partial_{t}\phi)^2
	+ f(\partial_{r}\phi)^2 
	+ \frac{1}{r^2}\Omega^{ij}\partial_{i}\phi\partial_{j}\phi
	+ m_{\phi}^2\phi^2\right]
\end{equation}
Hence, the contribution $\Delta M$ of the scalar field to the mass of
the background geometry is equal to the Hamiltonian of the field:
%
\begin{equation}
 \Delta M \equiv -\int_{r_{1}}^{L}T^t_t 4\pi r^2dr = H,
 	\label{eqn:DeltaM}
\end{equation}
where $H$ is given by (\ref{eqn:Hamiltonian}). 
When we consider the statistical mechanics of the hot quantized system, 
contributions of both vacuum polarization and thermal excitations must 
be taken into account. Thus, the contribution to the mass is given by
%
\begin{equation}
 \langle\Delta M\rangle = 
 	{\bf Tr}\left[e^{\beta_{\infty}(F-:H:)}\Delta M^{(ren)}\right],
\end{equation}
where $\Delta M^{(ren)}$ is an operator defined by the expression 
(\ref{eqn:DeltaM}) with $T^t_t$ replaced by the renormalized stress 
energy tensor $T_{\qquad t}^{(ren) t}$. From (\ref{eqn:DeltaM}), it is 
easy to show that 
%
\begin{equation}
 \Delta M^{(ren)} = :H: + \Delta M_{B},
\end{equation}
where $:H:$ is the normal-ordered Hamiltonian given by (\ref{eqn::H:}) 
and $\Delta M_{B}$ is the zero-point energy of the Boulware state defined 
by 
%
\begin{equation}
 \Delta M_{B} = -\int_{r_{1}}^{L}
 	\langle B|T_{\qquad t}^{(ren) t}|B\rangle 4\pi r^2dr.
\end{equation}
Hence, $\langle\Delta M\rangle$ can be decomposed into the
contribution of the thermal excitations and the contribution from the
zero-point energy:
%
\begin{equation}
 \langle\Delta M\rangle = U + \Delta M_{B},
 	\label{eqn:dM=dMB+U}
\end{equation}
where $U$ is given by (\ref{eqn:energy}) and equal to $\Delta M_{therm}$ 
defined in (\ref{eqn:Mtherm}).

Finally, we have shown that the gravitational mass appearing in the 
Einstein equation is the sum of the energy of the thermal excitation 
and the mass-energy of the Boulware state. Therefore, as shown in 
Sec.~\ref{sec:BW-model}, the wall contribution to the total gravitational 
mass is zero on shell ($T_{\infty}=T_{H}$) and the backreaction can 
be neglected. Here, we mention that the 
corresponding thermal state on shell is called a topped-up Boulware 
state~\cite{PVI}, and can be considered as a generalization to spacetimes 
not necessarily containing a black hole of the Hartle-Hawking 
state~\cite{HH}.


\section{Summary and discussion}
	\label{sec:discussion}

Attempts to provide a microscopic explanation of the
Bekenstein-Hawking entropy $S_{BH}$ initially stemmed from two quite
different directions. (See \cite{Frolov&Fursaev} for an up-to-date
review with full references.)

Gibbons and Hawking~\cite{Gibbons&Hawking} took the view that $S_{BH}$ 
is of topological origin, depending crucially on the presence of a
horizon. They showed that $S_{BH}$ emerges as a boundary contribution
to the geometrical part of the Euclidean action. (A non-extremal
horizon is represented by a regular point in the Euclidean sector, so
the presence of a horizon corresponds to the {\it absence} of an inner 
boundary in this sector.)

'tHooft~\cite{tHooft} sought the origin of $S_{BH}$ in the thermal
entropy of ambient quantum fields raised to the Hawking
temperature. He derived an expression which is indeed proportional to
the area, but with a diverging coefficient which has to be regulated
by interposing a ``brick wall'' just above the gravitational radius
and adjusting its altitude by hand to reproduce $S_{BH}$ with the
correct coefficient.

In addition, the brick wall model appears to have several problematical 
features---large thermal energy densities near the wall, producing a
substantial mass correction from thermal excitations---which have
raised questions about its self-consistency as a model in which
gravitational back-reaction is neglected.

We have shown that such caveats are seen to be unfounded once the
ground state of the model is identified correctly. Since there are no
horizons above the brick wall, the ground state is the Boulware state,
whose negative energy almost exactly neutralizes the positive energy
of the thermal excitations. 'tHooft's model is thus a perfectly
self-consistent description of a configuration which to outside
observers appears as a black hole but does not actually contain
horizons.

It is a fairly widely held opinion
(e.g. \cite{Callan&Wilczek,one-loop}) that the entropy contributed by
thermal excitations or entanglement is a one-loop correction to the
zero-loop (or ``classical'') Gibbons-Hawking contribution. The
viewpoint advocated in this paper appears (at least superficially)
quite different. We view these two entropy sources---(a) brick wall,
no horizon, strong thermal excitations near the wall, Boulware ground
state; and (b) black hole, horizon, weak (Hartle-Hawking) stress
-energy near the wall, Hartle-Hawking ground state---as ultimately
equivalent but mutually exclusive (complementary in the sense of Bohr) 
descriptions of what is externally virtually the same physical
situation. The near-vacuum experienced by free-falling observers near
the horizon is eccentrically but defensibly explainable, in terms of
the description (a), as a delicate cancellation between a large
thermal energy and an equally large and negative ground-state
energy---just as the Minkowski vacuum is explainable to a uniformly
accelerated observer as a thermal excitation above his negative-energy 
(Rindler) ground state. (This corresponds to setting $f(r)=r$ in the
$(1+1)$-dimensional example treated in Sec.~\ref{sec:B&HH-states}.)

That the entropy of thermal excitations can single-handedly account for 
$S_{BH}$ without cutoffs or other {\it ad hoc} adjustments can be
shown by a thermodynamical argument~\cite{PVI}. One considers the
reversible quasi-static contraction of a massive thin spherical shell
toward its gravitational radius. The exterior ground state is the
Boulware state, whose stress-energy diverges to large negative
values in the limit. To neutralize the resulting backreaction, the
exterior is filled with thermal radiation to produce a ``topped-up''
Boulware state (TUB) whose temperature equals the acceleration
temperature at the shell's radius. To maintain thermal equilibrium
(and hence applicability of the first law) the shell itself must be
raised to the same temperature. The first law of thermodynamics then
shows that the shell's entropy approaches $S_{BH}$ (in the
non-extremal case) for essentially arbitrary equations of state. Thus, 
the (shell $+$ TUB) configuration passes smoothly to a black hole $+$
Hartle-Hawking state in the limit.

It thus appears that one has two complementary descriptions, (a) and
(b), of physics near an event horizon, corresponding to different Fock 
representations, i.e., different definitions of positive frequency and 
ground state. The Bogoliubov transformation that links these
representations is known~\cite{Israel}. However, because of the
infinite number of field modes, the two ground states are unitarily
inequivalent~\cite{Umezawa}. This signals some kind 
of phase transition (formation of a condensate) in the passage
between description (a), which explains $S_{BH}$ as a thermal effect,
and description (b), which explains it as geometry. 
We know that a condensation actually does occur at this point; it is
more usually called gravitational collapse.

It will be interesting to explore the deeper implications of these
connections.

\vskip 1cm

\centerline{\bf Acknowledgments}

One of us (S. M.) thanks Professor H. Kodama for his continuing 
encouragement. 
The other (W. I.) would like to thank Professor H. Kodama, T. Nakamura 
and K. Tomita and their associates for the friendly and stimulating
atmosphere they have provided at the Yukawa Institute for Theoretical
Physics. He is very grateful to Professor H. Sato and M. Fukugita for
their hospitality and many kindnesses. He would also like to
acknowledge stimulating discussion with Professor M. Sasaki and his
group at Osaka University, and with Professor A. Hosoya and
H. Ishihara and their group at the Tokyo Institute of Technology. 
S. M. was supported by JSPS Research Fellowships for Young Scientists, 
and this work was supported partially by the Grant-in-Aid for
Scientific Research Fund (No. 9809228).
W. I. was supported by the Ministry of Education, Science, Sports
and Culture, and partly by the Canadian Institute for Advanced
Research and NSERC of Canada.


\appendix


\section{On-shell brick wall model}
	\label{app:on-shell}

When we performed the differentiation with respect to 
$\beta_{\infty}$ to obtain the total energy and the entropy, the 
surface gravity $\kappa_{0}$ of the black hole and the inverse 
temperature $\beta_{\infty}$ of gas on the black hole background were
considered as independent quantities. Since in equilibrium these quantities 
are related by $\beta_{\infty}^{-1}=\kappa_0/2\pi$, we have imposed this
relation, which we call the on-shell condition,  after the differentiation. 
In fact, we have shown that the wall contribution to gravitational energy 
is zero and the backreaction can be neglected, if and only if the 
on-shell condition is satisfied.

On the other hand, in the so-called on-shell 
method~\cite{on-shell,Belgiorno&Martellini}, the on-shell condition is  
implemented before the differentiation. 
Now let us investigate what we might call an on-shell brick wall model. 
With the on-shell condition, the wall contribution to the free energy 
of the scalar field considered in Sec.~\ref{sec:BW-model2} is 
calculated as 
%
\begin{equation}
 F_{wall}^{(on-shell)} = 
 	-\frac{A}{4}\frac{\beta_{\infty}^{-1}}{360\pi}\frac{1}{\alpha^2}.
\end{equation}
If we define total energy and entropy in the on-shell method by 
%
\begin{eqnarray}
 U_{wall}^{(on-shell)} & \equiv & 
 	\frac{\partial}{\partial\beta_{\infty}}
 	\left(\beta_{\infty} F_{wall}^{(on-shell)}\right),\nonumber\\
 S_{wall}^{(on-shell)} & \equiv & 
 	\beta_{\infty}^2\frac{\partial}{\partial\beta_{\infty}}
 	F_{wall}^{(on-shell)},
\end{eqnarray}
then these quantities can be calculated as 
%
\begin{eqnarray}
 U_{wall}^{(on-shell)} & = & 0,\nonumber\\
 S_{wall}^{(on-shell)} & = & 
 	\frac{A}{4}\frac{1}{360\pi}\frac{1}{\alpha^2}
	=\frac{1}{4}S_{wall},
\end{eqnarray}
where $S_{wall}$ is the wall contribution (\ref{eqn:Swall2}) to 
entropy of the scalar field with $T_{\infty}=T_{H}$.

It is notable that the total energy $U_{wall}^{(on-shell)}$ in the 
on-shell method is zero irrespective of the value of the cutoff 
$\alpha$. However, $S_{wall}^{(on-shell)}$ is always smaller than 
$S_{wall}$. It is because some physical degrees of freedom are frozen 
by imposing the on-shell condition before the differentiation. Thus, we 
might miss the physical degrees of freedom in the on-shell method.


\end{document}